\documentclass[preprint,12pt]{aastex}
\usepackage{natbib}
\usepackage{graphicx}
\usepackage{latexsym}

\begin{document}

\title{The Mid-Infrared Instrument for the James Webb Space Telescope, IX: 
Predicted Sensitivity}

\author{Alistair Glasse\altaffilmark{1}, G. H. Rieke\altaffilmark{2}, E. 
Bauwens\altaffilmark{3}, Macarena Garc\'ia-Mar\'in\altaffilmark{4}, M. E.  
Ressler\altaffilmark{5}, Steffen Rost\altaffilmark{4}, T. V. Tikkanen\altaffilmark{6}, 
B. Vandenbussche\altaffilmark{3}, G. S. Wright\altaffilmark{1}}

\altaffiltext{1}{UK Astronomy Technology Centre, Royal Observatory, Edinburgh, Blackford Hill, Edinburgh
EH9 3HJ, UK}
\altaffiltext{2}{ Steward Observatory, 933 N. Cherry Ave, University of Arizona, Tucson, AZ 85721, USA}
\altaffiltext{3}{Institute of Astronomy KU Leuven, Celestijnenlaan 200D,3001 Leuven, Belgium}
\altaffiltext{4}{I. Physikalisches Institut, Universität zu Köln, Zülpicher Str. 77,  50937 Köln, Germany }
\altaffiltext{5}{Jet Propulsion Laboratory, California Institute of Technology, 4800 Oak Grove Dr. Pasadena, CA 91109, USA}
\altaffiltext{6}{Department of Physics and Astronomy, Univ. of Leicester, University Road, Leicester, LE1 7RH, UK}

\begin{abstract}

We present an estimate of the performance that will be 
achieved during on-orbit operations of the JWST Mid Infrared Instrument, 
MIRI. The efficiency of the main imager and spectrometer systems in 
detecting photons from an astronomical target are presented, based on 
measurements at sub-system and instrument level testing, with the end-to-end 
transmission budget discussed in some detail. The brightest target fluxes 
that can be measured without saturating the detectors are provided. The 
sensitivity for long duration observations of faint sources is presented in 
terms of the target flux required to achieve a signal to noise ratio of 10 
after a 10,000 second observation. The 
algorithms used in the sensitivity model are presented, including the 
understanding gained during testing of the MIRI Flight Model and flight-like 
detectors. 

\end{abstract}

Keywords: Astronomical Instrumentation

\section{Introduction}
MIRI has been designed (Wright et al., 2014, hereafter Paper II) to detect as much as possible 
of the signal from the astronomical target 
collected by the JWST primary aperture, and so the 
measurement noise is limited to the statistical fluctuations on the total 
signal (target plus background), commonly referred to as shot noise. 
In practice, the non-ideal behaviour of optics and detector systems mean 
that this goal can be closely approached, but never reached: not all light 
collected by the telescope is detected and at the short wavelength end of 
MIRI's spectral range detector dark current becomes significant, most 
notably for the medium resolution spectrometer. The contributions of the 
opto-electronic sub-systems to MIRI's sensitivity have been carefully 
characterized during the instrument build, by direct measurement at 
component level combined with system level analysis, and culminating in 
extended cryogenic testing of the fully assembled MIRI flight model at the 
Rutherford Appleton Laboratory (RAL) during 2011, in a flight-like radiative 
environment using a well characterized radiometric source (the MIRI 
Telescope Simulator, MTS,discussed in Paper II). Our understanding of the detector performance 
(Rieke et al, 2014, hereafter Paper VII) has benefitted from an extensive 
programme of performance testing conducted using flight-like detectors at 
JPL. 

These performance metrics were brought together in a sensitivity model that 
was formulated early on in the project (Swinyard et al., 2004). This has 
since been extended and refined to track the impact of developments to the 
observatory design on MIRI's expected on-orbit performance and to include 
the knowledge gained during the MIRI build and test phases. It includes 
performance estimates at all wavelengths and optical configurations. 

In this paper we present and justify the parameters and algorithms that 
comprise the sensitivity model, and present its predictions for three of MIRI's four
major operational configurations; the imager, the low resolution 
spectrometer (LRS) and the medium resolution integral field spectrometer 
(MRS). Coronagraphy is not included because its performance is more 
dependent on contrast and rejection than on throughput and sampling as dealt with 
in this paper. The predicted coronagraph performance is discussed in Boccaletti et 
al., (2014, Paper V). 

The results presented here should be 
interpreted as the limiting sensitivities that could be achieved during 
long duration (hour or more) staring observations made at a single 
spacecraft pointing. They exclude inefficiencies due to the operational 
overheads described in Gordon et al., (2014), Paper X, such as target acquisition, 
small angle manoeuvres to allow efficient background subtraction, time spent 
moving MIRI mechanisms or taking calibration observations. 

\section{Model Components}
\subsection{The JWST Observatory}
The massive gains in sensitivity that MIRI will achieve, compared to current 
and planned mid-infrared instruments, are primarily due to the large (25 
m$^{\mathrm{2}})$ collecting area and the cold (40 K) radiative environment 
provided by the JWST. When compared to ground-based observatories, the 
absence of atmospheric absorption bands and thermal emission from both the atmospheric and telescope makes JWST 
more than competitive even with the 30- and 40-m class telescopes now being planned.

At the short wavelength end of MIRI's spectral range around 5 $\mu$m, the remaining 
radiative background signal is dominated by emission from the zodiacal dust 
concentrated in the ecliptic plane; longward of 17 $\mu$m or so, 
straylight from the observatory sunshield and the thermal emission of the 
telescope optics become dominant. 

\begin{figure}[htbp]
\centerline{\includegraphics[width=5.0in]{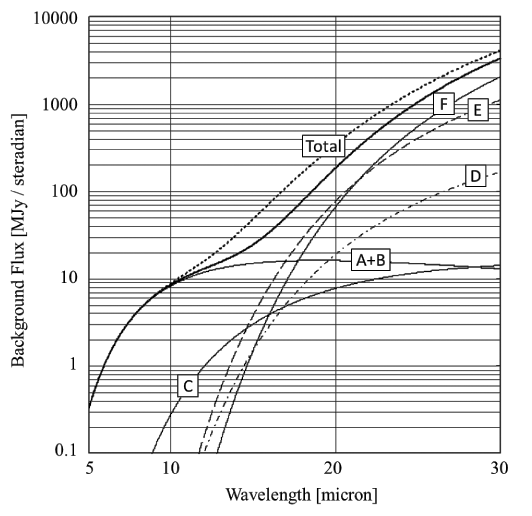}}

\caption{Background emission spectra used in MIRI sensitivity modelling. }
\label{fig1}
\end{figure}

For the purposes of sensitivity modelling, we represent the backgrounds by the sum of six 
grey body emission spectra, whose emissivities and effective temperatures 
are listed in \tablename~\ref{tab1}, with spectral energy 
distributions plotted in \figurename~\ref{fig1}, where the 
background spectrum is seen to rise steeply across MIRI's spectral range. 

Components A and B in \tablename~\ref{tab1} are fits to the 
scattered and emissive components of the zodiacal dust spectrum towards the 
celestial north pole (Wright, 1998), scaled by a factor of 1.2 to be 
representative of typical pointing scenarios. Components C to F are derived 
from a fit to a detailed straylight model of the observatory background 
(Lightsey and Wei, 2012), such that individual terms should not be 
identified as being physically representative of any specific observatory 
sub-system. This particular model gives a total of 188 MJy/sr at 20 $\mu$m, just 
under the second-level requirement of 200 MJy/sr.

To include the thermal background from the telescope, components 
A to F are summed together to give the background spectrum that is plotted 
as the thick solid line labelled `TOTAL' in Figure~\ref{fig1}. As a contingency, we also show the 
result of doubling the current estimates of facility emission by adding an additional component 
(G in \tablename~\ref{tab1}) with the effect of increasing the background flux at $\lambda $ 
$=$ 20 $\mu $m to 350 MJy steradian$^{\mathrm{-1}}$. 

In interpreting \figurename~\ref{fig1}, it may be helpful to note 
that for the MIRI imager pixel scale of 0.11 arcsecond, the useable area 
of the JWST entrance pupil ($A_{tel} =$ 25.03 m$^{\mathrm{2}})$ and the 
nominal transmission of the telescope optics up to the MIRI entrance focal 
plane ($\tau_{tel} =$ 0.88 at the start of the mission), a flux of 0.4 
MJy sterad$^{\mathrm{-1}}$ at $\lambda =$ 5 $\mu $m corresponds to 7.5 
photon sec$^{\mathrm{-1}}$ pixel$^{\mathrm{-1}}$ in a 1 micron pass-band. At 
$\lambda =$ 20 $\mu $m, a flux of 188 MJy sterad$^{\mathrm{-1}}$ equates 
to 890 photon sec$^{\mathrm{-1}}$ pixel$^{\mathrm{-1}} \quad \mu 
$m$^{\mathrm{-1}}$.

\subsection{Photon Conversion Efficiency}
The Photon Conversion Efficiency (PCE) is defined as the fraction of photons 
crossing the MIRI focal plane within the science field of view that are 
detected (and so contribute to the measurement). It appears in the following 
formula describing the photocurrent `$i$' generated in a detector pixel by a 
photon flux entering the telescope, `$P(\lambda )$' (with units of photon 
steradian$^{\mathrm{-1}}$ micron$^{\mathrm{-1}})$. 

\begin{equation}
\label{eq1}
i = \Omega_{pix}\tau_{tel}\tau_{EOL}\int_{\Delta \lambda} P \left( \lambda \right) \tau_{\lambda}\eta_{\lambda}d\lambda
\end{equation}

Here, the PCE is written as the product of the optical transmission of all 
elements from the focal plane to the detector `$\tau_{\mathrm{\lambda 
}}$', with the detector quantum efficiency `$\eta_{\mathrm{\lambda }}$' 
(measured in detected electrons per incident photon). The factor `$\Omega 
_{pix}$', which represents the solid angle field of view of a single 
pixel, is combined with two additional wavelength independent transmission 
terms; `$\tau_{tel}$' (set equal to 0.88), describes the transmission of 
the clean telescope optics at the start of the mission, and `$\tau 
_{EOL}$' (set equal to 0.80) is then used to account for the loss in 
transmission of all elements in the optical train up to the `End Of Life' of 
the nominal 5 year mission. 

We determined the `beginning of life' PCE in two independent ways, with the 
results for the imager shown in Figure 2 and for 
the MRS shown in \figurename~\ref{fig3}. 

\begin{figure}[htbp]
\centerline{\includegraphics[width=5.0in]{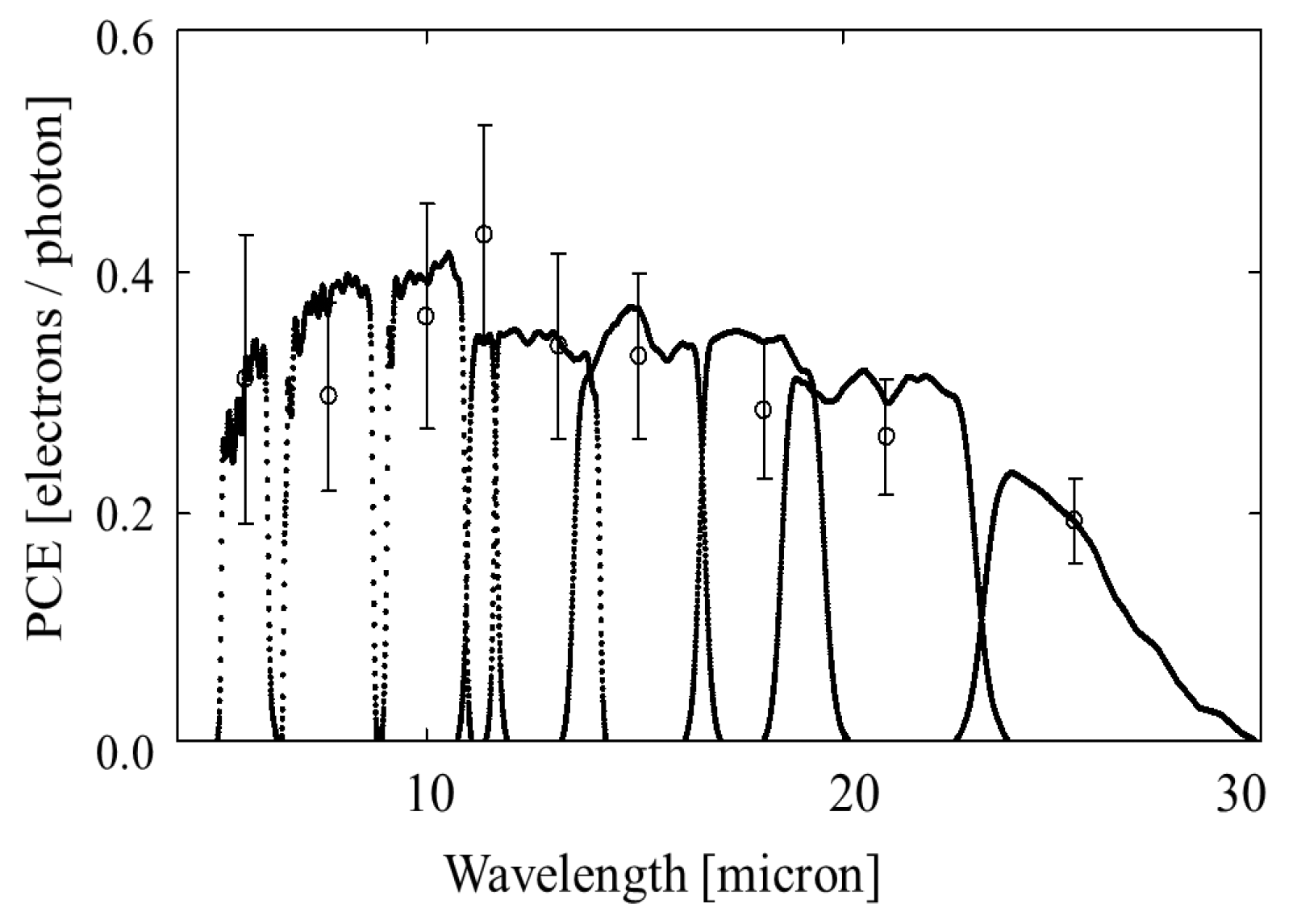}}
\caption{Photon Conversion Efficiency (detected electrons per photon crossing the MIRI Imager entrance focal plane).}
\label{fig2}
\end{figure}

\begin{figure}[htbp]
\centerline{\includegraphics[width=5.0in]{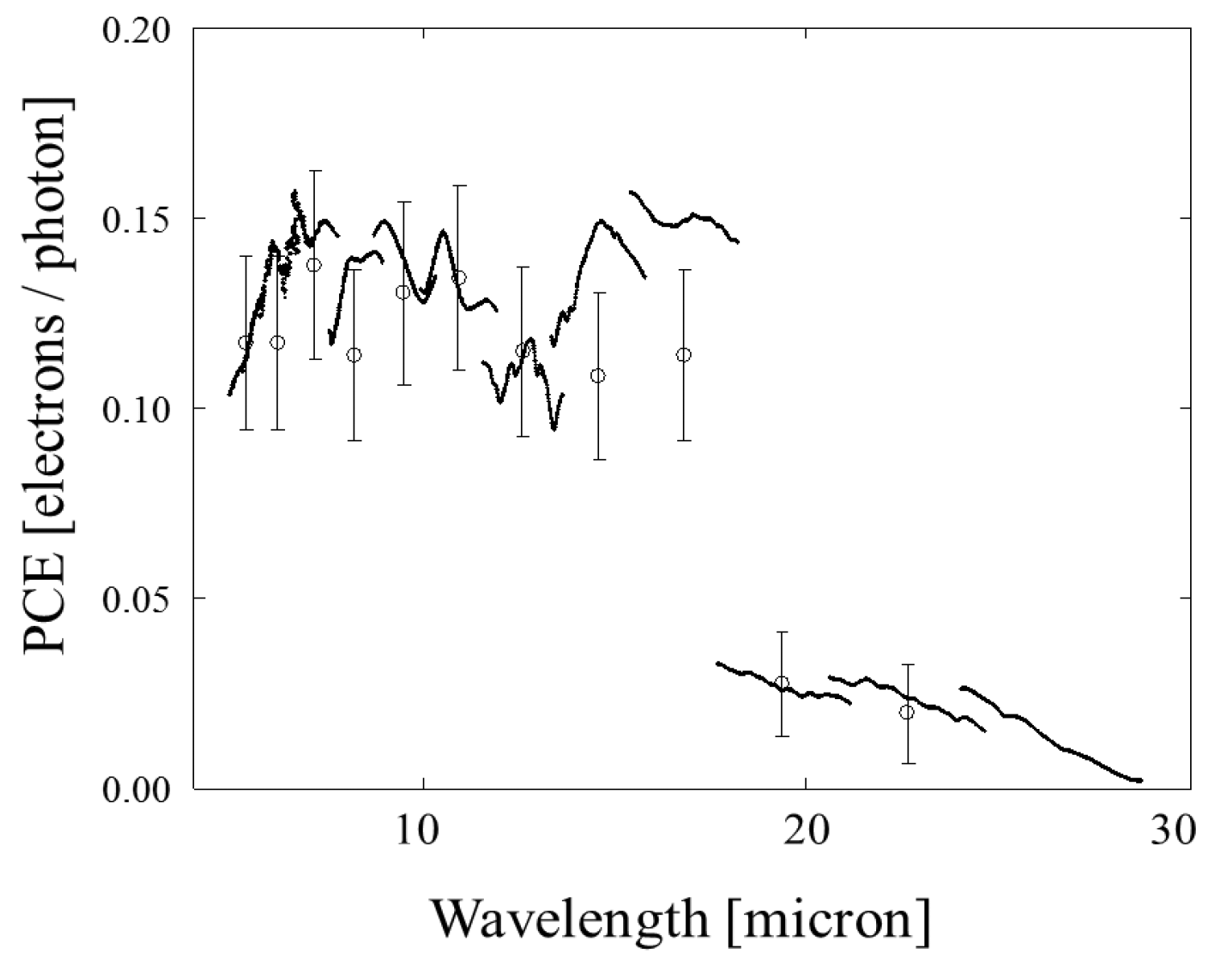}}
\caption{Photon Conversion Efficiency (units of electron photon-1) of the MIRI Medium Resolution Spectrometer. }
\label{fig3}
\end{figure}

For the first method, the measured and estimated transmission efficiencies 
due to all elements in the optical train were combined with the partly 
measured, partly modelled, detector quantum efficiencies given in Paper VII, to 
generate `bottom-up' wavelength dependent PCE profiles for the imager, LRS 
and MRS. These profiles are plotted as the solid lines in 
 \figurename~\ref{fig2} and \figurename~\ref{fig3}.

In the case of the imager, the wavelength dependent transmissions of the 
band selection filters were measured at the 7 K operating temperature by the 
filter manufacturers (University of Reading and Spectrogon). The seven gold 
coated mirrors in the optical train from the entrance focal plane were then 
allocated a reflectivity of 0.98 per surface (consistent with cryogenic 
measurements made on similar mirrors). An additional transmission factor of 
0.8 was applied to the end-to-end transmission to allow for contamination up 
to the start of the mission (i.e. independent of the end-of-life factor 
`$\tau_{EOL}$' referred to above).

For the MRS (whose optical train is described in Wells et al., 2014, 
hereafter Paper VI), the transmissions of the dichroic filters were again 
derived from the manufacturers (University of Reading) cryogenic 
measurements. The same reflectivity figure of 0.98 was used for the mirrors, 
where we note that the number of mirrors in each of MRS Channels 1 to 4 is 
26, 25, 24 and 29 respectively. 

As discussed in Paper VI, the optical design of the MRS was tailored to 
minimize losses due to diffraction, primarily by oversizing optical elements 
lying between the integral field units and the detectors. However, some 
losses were expected to remain so modelling (using the `GLAD' software 
package) was used to derive effective transmissions `$\tau 
_{MRS\mathunderscore diff}$' at the short and long wavelength ends of each 
MRS channel, accounting for diffraction losses. These transmissions are 
tabulated in \tablename~\ref{tab2}. For use in the sensitivity 
model, the value of $\tau_{MRS\mathunderscore diff}$ at a specific 
wavelength was derived by linear interpolation. 

The efficiency of the diffraction gratings was set to be 0.6, based on 
modelling (using PCGrate) for all gratings in Channels 1, 2 and 3. For 
Channel 4, lower than expected efficiency was measured during testing at RAL (see 
Paper VI). This was accounted for in the sensitivity model by setting the 
grating efficiencies in this case equal to 0.17. 

The product of the above transmissions and efficiencies, with the same 
additional end-to-end factor of 0.8 as used for the imager to account for 
beginning of life contamination, is then plotted as the solid line in 
\figurename~\ref{fig3}. 

A second method for estimating the PCE independently of the sub-system 
budgets described above involved direct measurement of the signal during 
observations of the extended source of the MIRI Telescope Simulator (MTS) 
during Flight Model testing at RAL in 2011. An end-to-end model of the photometric output of the MTS (called MTSSim), 
described in Paper II, was then used to provide an absolute flux 
calibration of these measurements, with a value of 5.5 electrons per data 
number assumed for the electronic gain of the detector system (Paper VIII). 
These band averaged results are plotted in 
Figure 2 and \figurename~\ref{fig3} as 
points with error bars. We note that it was necessary to scale the PCE 
values predicted by MTSSim (version 2.2) by a factor of 0.55 for both the 
imager and MRS to obtain the reasonable agreement with the MIRI 
sub-system derived profiles seen in Figures 2 and 3. This factor was 
ascribed in Paper II to systematic errors in MTSSim; we note 
here that one alternative, namely that the MIRI sub-system determined PCEs 
presented here are all systematically too low by a factor of 0.55, is felt 
to be unlikely; it is difficult to identify a single component common to 
both the MRS and the imager that could credibly have a transmission that 
is almost a factor of 2 higher than the values described above. 

\subsection{Encircled Energy}
Not all of the signal detected from a point-like astronomical target will 
contribute to the photometric measurement. The signal is distributed over 
many pixels due to image broadening by the telescope and instrument optics 
and by scattering in the detector itself (Paper VII). This broadening, which 
is quantified as the instrument Point Spread Function (PSF), forces us to 
make a trade, which maximizes the fraction of the total integrated signal 
from the target that is sampled, whilst minimizing the number of pixels 
contributing noise. 

If we consider the target as a point source of total integrated flux 
`$F_{point\mathunderscore tgt}$' (Jansky), then the photon flux (in units of 
photon sec$^{-1}$ micron$^{-1}$ arcsec$^{-2}$) contributing to the 
measured signal can be calculated in terms of the fraction `$f_{phot}$' of the 
total falling within a circular photometric aperture of radius 
`$r_{phot}$', containing `$N_{phot}$' pixels, each of which views a solid 
angle on the sky equal to `$\Omega_{pix}$'.
\begin{equation}
\label{eq1}
P_{point\_ tgt} = \frac{F_{point\_tgt}f_{phot}}{\Omega_{pix}N_{phot}}\frac{A_{tel}}{h\lambda}
\end{equation}
Testing of the MIRI Flight Model has shown that the images delivered to the 
detector by the MIRI optics at wavelengths longward of $\lambda =$ 7 
$\mu $m are near-diffraction limited, allowing us to make a straight-forward 
definition of the photometric aperture radius, $r_{phot}$ as being equal to 
that of the first dark Airy ring. For the JWST's non-circular pupil we 
approximate this as,
\begin{equation}
\label{eq2}
r_{phot} = 0 \farcs42 \left(\frac{\lambda}{10 \mu m}\right)
\end{equation}
We choose to express the encircled energy fraction, $f_{phot}$ as the product 
of two factors; `$f_{opt}$', which describes the effects of image broadening 
due to misalignment (primarily defocus) between MIRI and the JWST, and 
`$f_{det}$' which accounts for the flux lost outside the photometric aperture 
due to scattering in the detector. 

The physical origin and impact on the image quality of scattering within the 
detector are described in Paper VII. For inclusion in the sensitivity model 
we represent this effect with the function, 
\begin{equation}
\label{eq3}
f_{det}=1-{a_{det}e}^{-\left( \left[ \tau_{1}\thinspace +\thinspace \tau 
_{2} \right]\left( \frac{\lambda }{7\thinspace \mu m} \right)^{2} \right)}
\end{equation}
Here the factor `$\tau_{\mathrm{1}}$' (equal to 0.36) is the optical 
depth at $\lambda \quad =$ 7 $\mu $m for absorption by a single pass through 
the 35 $\mu $m thick active layer of the MIRI detector for light at normal 
incidence, as defined in Paper VII. In the simple case when factors 
$a_{det} \quad =$ 1 and $\tau_{\mathrm{1}} \quad =$ 0, $f_{det}$ would then model 
the case where all light that is not absorbed within the active layer on 
the way in is lost from the photometric aperture. The factors $\tau 
_{\mathrm{2}}$ and $a_{det}$ then refine this picture to better fit two 
features of the scattering as measured during flight model testing. 

First, as described in Bouchet et al. (2014, Paper III), the measured dimensions of the 
imager PSF are close to the expectations of diffraction limited performance. 
Detailed optical modelling then allowed an upper limit to be placed on the 
fraction of light scattered to large radii (i.e outside the photometric 
aperture) equal to 15 {\%} for the F560W filter. This constraint is 
satisfied by setting $a_{det}$ equal to 0.32. We might interpret $a_{det}$ as 
a measure of the wavelength independent component of the scattered light.

The second observed feature of the detector scattering, embodied in factor 
`$\tau_{\mathrm{2}}$' comes from flood illuminated images measured during 
the RAL test campaign. It was found that the straylight falling on regions 
of the imager occulted by the focal plane mask (Paper III) could 
be well modelled by the expected diffraction limited PSF at wavelengths 
longward of 10 $\mu $m, but that for the F560W and F770W filters an 
additional component was needed, which was ascribed to detector scattering. 
The amplitude of this component (which equates to 1 -- $f_{det})$ was found 
to be a factor of 2 larger for the F560W filter than for F770W. Matching 
this ratio for $\tau_{\mathrm{1}} \quad =$ 0.36 requires that $\tau 
_{\mathrm{2}} \quad =$ 0.85. 

The functional similarity between $\tau_{\mathrm{1}}$ and $\tau 
_{\mathrm{2}}$ should not be interpreted as implying a similar physical 
origin. As discussed in Paper VII, the details of the scattering process are 
complicated; our aim in Equation (\ref{eq3}) is simply to provide a representation 
of its coarse behaviour suitable for inclusion in the sensitivity model. 

For ideal optical performance (a Strehl ratio of 100 {\%}) and perfect 
optical alignment between MIRI and the telescope, $r_{phot}$ will encircle a 
fraction $f_{optA} = $70 {\%} of the flux for a point-like target. Optical 
modelling of MIRI on the JWST has allowed us to derive values for 
$f_{opt}$ (with an estimated error of $\pm$1 {\%}) under a range of 
non-ideal conditions of defocus and pupil shear misalignment. The `worst 
case' `$f_{optB}$' occurs where the optical train (MIRI plus telescope) 
introduces a wavefront error equivalent to an 85 {\%} Strehl ratio, on top 
of which there is 5 mm of defocus and 2 {\%} pupil decentre between the 
focal plane delivered by the JWST and that accepted by MIRI. This function 
is well described by, 
\begin{equation}
\label{eq4}
f_{optB}=0.68\thinspace \left( 1-\left( \frac{2.47\mu m}{\lambda } 
\right)^{1.8} \right)
\end{equation}
The limits on $f_{phot}$ for the imager set by these best and worst case 
bounds on $f_{opt}$ are plotted as the dotted lines in 
\figurename~\ref{fig5}. 

Measurements of the co-alignment between the MIRI Flight Model and the 
JWST-ISIM instrument module showed that the defocus and pupil decentre were 
both consistent with the case of `perfect' optical alignment, whilst the 
measured image quality suggested that the delivered Strehl ratio at the 
detector was in excess of 95 {\%}. We therefore chose to retire a 
substantial fraction of the risk associated with these optical effects by 
adopting a weighted sum of the limiting cases, 
\begin{equation}
\label{eq5}
f_{opt}=0.75\thinspace f_{optA}+\thinspace 0.25\thinspace f_{optB}
\end{equation}
It is this function that was used to generate the encircled energy fraction 
$f_{phot} = \quad f_{det} \times f_{opt}$ in the sensitivity model and that 
is plotted for the imager and LRS in \figurename~\ref{fig5}. 

We must adapt this approach for the low resolution spectrometer (the LRS, 
described in Paper IV), and the medium resolution spectrometer (the MRS, 
described in Paper VI) as a result of the differences in the dimensions of 
their respective photometric apertures. 

For the LRS, the aperture is assumed to be rectangular, with dimensions 
equal to the slit width (0.51 arcsecond, or 4.6 pixels) in the across-slit 
direction and by $r_{phot}$ in the along slit direction. We then need to 
modify the form of $f_{phot}$ to account for the change in encircled energy 
fraction due to the non-circular aperture. Numerical integration of the 
nominal PSF was used to determine a correction factor, equal to the ratio of 
the energy within the rectangular LRS aperture to the circular aperture 
defined by $r_{phot}$, normalized at $\lambda =$ 10 $\mu $m. This was 
found to be well fitted (maximum deviations \textless 0.01) by a linear 
function, which generates correction factors for $f_{phot}$ that vary from 
1.13 at $\lambda =$ 5 $\mu $m to 0.85 at $\lambda =$ 15 $\mu $m. 

For the MRS, we note that the photometric aperture is sampled in one 
direction (parallel to the integral field slices described in Paper VI) by 
the detector pixels and in the orthogonal direction by the slices 
themselves. The detector scattering described above will therefore only 
apply in the along slice direction of the photometric aperture. It will have 
a different effect in the dispersion direction, where it will cause light to be 
scattered outside the nominal spectral resolution element. 

The magnitude of the impact on the MRS' sensitivity is then determined by 
the size of the photometric aperture in the along-slice direction and the 
width of the spectral resolution element, both measured in pixels. The field 
of view of a single MRS pixel is 0.20 arcsecond for $\lambda $ \textless 12 
$\mu $m (see Paper VI), a factor of 1.8 larger than that of the imager. In 
the spectral direction, the half-width of the spectral resolution element is 
roughly 0.9 pixels across this spectral range (as shown in 
\figurename~\ref{fig7}), a factor of 2.4 smaller than the 
diameter of the imager photometric aperture at $\lambda =$ 7 $\mu $m. We 
therefore expect detector scattering to result in greater loss of signal for 
the MRS compared to the imager. 

In the absence of a good understanding of the spatial distribution of 
detector scattering we were unable to derive an accurate quantitative 
correction to $f_{phot}$ for the MRS. Instead, inspection of the measured MRS 
PSF (specifically, Figure 16 in Paper VI) suggests that we should increase 
the linear size of the photometric aperture by 50 {\%} in the along-slice 
direction to accommodate the observed extension, while leaving its 
across-slice dimension unchanged. For the sensitivity model, this is 
implemented by increasing $r_{phot}$ by a factor of $\surd $1.5 $=$ 1.22 for 
the MRS. We note that this factor may prove to be conservative at long 
wavelengths ($\lambda $ \textgreater 10 $\mu $m) where the along-slice 
extension should be reduced (irrespective of whether it is due to scattering 
in the detector or due to optical scattering at surfaces in the MRS). 

\subsection{Detector Performance}
As described in more detail in Papers VII and VIII, the electrons generated in each 
pixel are used to charge a $\sim$34 fF capacitor (with a 250,000 
electron storage capacity in MIRI's case), which is sampled (`read') 
multiple times at a fixed interval `$t_{frame}$'. The photocurrent is then 
determined by finding the slope that best fits these samples, with a 
measurement error that includes terms due to the shot noise on the charge, 
the read noise associated with each sample, and the statistical effects of 
sampling up the ramp. 

Analytical expressions that describe the signal and noise have been derived 
for example by Herter (1989), whilst Garnett and Forrest (1993) describe the 
terms that account for the sampling errors for the slope-fitting sampling 
scheme used by MIRI. We have combined these to describe the signal and noise 
per pixel in a single integration in which the photocharge `$i$', is integrated 
for a duration `$t_{int}$',
\begin{equation}
\label{eq6}
S_{int}=i_{sig}t_{int}
\end{equation}

\begin{equation}
\label{eq6}
N_{int}^2 = k1~ \left( i_{sig}+i_{bgd}\right)t_{int}+k2~i_{dark}~t_{int}+k3~R_N^2
\end{equation}

\noindent
Here, we separate the noise into three terms, which are related to: 1.) the total 
incident photon flux, 2.) the detector dark current (charge generated within the 
pixel in the absence of any illumination), and 3.) the read noise per pixel on a 
single frame `$R_{N}$'. The constant factors in each term include the 
statistical effect of sampling the charge with `$n_{read}$' equispaced pixel 
reads (Garnett and Forrest, 1993) as set out below, 

\begin{equation}
\label{eq9}
k1 = k_{exc}\frac{6}{5}\left(\frac{n_{read}^2+1}{n_{read}^2-1}\right)
\end{equation}
\[
k2=1
\]
\[
k3=k_{RNobs}\frac{12\thinspace n_{read}}{n_{read}^{2}-1}
\]
We see that term $k$1 describes the factor by which the measured noise exceeds 
the theoretical shot noise limit under conditions where the signal from the 
astronomical target or the background dominate. The factor `$k_{exc}$' 
replaces the more familiar product of gain and gain dispersion ($\beta $G); 
as described in Paper VII, the detector model is consistent with values 
for $\beta $ and G both being close to unity, but the detectors do exhibit 
noise which has been measured (Paper VII) during testing of the MIRI 
detectors to be significantly above the theoretical shot noise. For the 
sensitivity model we use a value of $k_{exc} \quad =$ 1.3. 

Based on detector measurements (Paper VIII), the noise associated with the 
dark current has been shown to follow Poissonian statistics (i.e behave as 
shot noise), leading us to set factor \textit{k2} equal to unity. Dark current values 
of 0.12, 0.03 and 0.10 el pixel$^{\mathrm{-1}}$ sec$^{\mathrm{-1}}$ are 
adopted for the imager, short wavelength MRS and long wavelength MRS 
detectors, respectively.

For faint sources and low backgrounds, the factor \textit{k}3 combines the ideal 
statistical reduction in the effective read noise compared to `$R_{N}$' with 
an additional factor `$k_{RNobs}$' (set equal to 0.815, Ressler, private 
communication), which reflects the measured factor by which the system noise 
falls short of this ideal behaviour. The value of $R_{N}$ is taken from Paper 
VIII as 32.6 electrons when the detector is operated in FAST mode. For SLOW 
mode, this is reduced by a factor of $\surd $8 to account for the 8 samples 
per pixel that contribute to each frame. 

The per pixel photocurrents from the target and background ($i_{sig}$ and 
$i_{bgd}$ in Equation 8) are then derived by application of Equation 1 to the 
relevant illumination source. For the special case of the (continuum) 
background spectrum seen by the LRS, the solid angle subtended by a single 
imager pixel (0.11 x 0.11 arcsecond$^{\mathrm{2}})$ is scaled by a factor of 
4.6 to account for the fact that each pixel in the LRS spectrum will see the 
co-added background from all (4.6) pixels across the slit. The wavelength 
dependent fields of view of the four MRS spectral channels are specified in 
Paper VI. 

There are several other factors that contribute to the sensitivity model. 
These include an efficiency factor that accounts for measurements that are 
lost due to the impact of cosmic rays. The effect of bad pixels and cosmic 
rays is currently folded in with the estimation of the effective exposure 
time `$t_{eff}$' that contributes to the integration for an exposure of 
duration `$t_{exp}$', 
\begin{equation}
t_{eff}=f_{goodEOL}t_{exp}\left( 1-R_{\gamma }t_{int}-\raise0.7ex\hbox{$1$} 
\!\mathord{\left/ {\vphantom {1 
n_{read}}}\right.\kern-\nulldelimiterspace}\!\lower0.7ex\hbox{$n_{read}$} 
\right)
\end{equation}

\noindent
Here, `$R_{\gamma }$' ($=$ 4.7 x 10$^{\mathrm{-4}}$ sec$^{\mathrm{-1}})$ is 
the estimated rate at which pixels will be disrupted by cosmic ray impacts 
on orbit, and `$f_{goodEOL}$' ($=$ 95 {\%}) is the fraction of pixels 
expected to be remaining in an operable state at the end of mission life. 
The final term uses `$n_{read}$', the number of up-the-ramp reads in an 
integration to account for the time lost between the reset and the first 
frame of the next integration. 

It should be noted that this approach to cosmic ray events is somewhat 
conservative, since it assumes that the entire integration ramp is lost. In 
practice, the slopes before and after an event can be recovered using 
algorithms developed by the MIRI team (Gordon et al. 2014, Paper X). 

\section{Model Predictions}
\subsection{Bright Source Limits}
We express the flux of a bright astronomical target as, 
\begin{equation}
\label{eq5}
F_{bright}=\frac{i_{bright}-i_{bgd}}{f_{br\mathunderscore pix}\thinspace g}
\end{equation}
Here, `$i_{bright}$' is the maximum acceptable photocurrent from a bright 
astronomical target, defined as that which will fill the brightest pixel in 
the image to 150,000 electrons (60 {\%} of the storage capacity quoted 
above) in the 5.6 seconds taken for the shortest useful full frame 
integration (comprising a reset and two `FAST' mode reads, as defined in 
Paper VIII). This equates to 26,800 el sec$^{\mathrm{-1}}$ 
pixel$^{\mathrm{-1}}$. 

The factor `$i_{bgd}$' accounts for the photocurrent generated by the 
background, which for the worst case of the F2550W filter and the `high 
background' case defined above is estimated to be around 5,100 el 
sec$^{\mathrm{-1}}$.
The factor `$g$' is then the end-to-end gain from astronomical target to 
photocurrent (with units of el sec$^{\mathrm{-1}}$ Jansky$^{\mathrm{-1}}$). This is 
calculated as a by-product of the sensitivity model using the PCE curves 
derived above and accounting for the relevant encircled energy and sampling 
factors. The target spectral shape is modelled as a 5,000 K 
black body.

The factor `$f_{br\mathunderscore pix}$' in Equation 11 describes the 
fraction of the total signal from a point target that will fall in the 
brightest pixel. For the imager, based on analysis of model PSFs, we set 
$f_{br\mathunderscore pix\mathunderscore imager} \quad =$ 0.13 for $\lambda $ 
$\le $ 8 $\mu $m. At longer wavelengths, to allow for image dilution by 
diffractive broadening we use, 
\begin{equation}
{\begin{array}{*{20}c}
f_{br\mathunderscore pix\mathunderscore imager}=0.13\thinspace \left( 
\raise0.7ex\hbox{${8\thinspace \mu m}$} \!\mathord{\left/ {\vphantom 
{{8\thinspace \mu m} \lambda 
}}\right.\kern-\nulldelimiterspace}\!\lower0.7ex\hbox{$\lambda $} 
\right)^{2}\\
\thinspace \thinspace \thinspace \thinspace \thinspace \thinspace \\
\end{array} }
\end{equation}

The resulting bright source limits (calculated for the `High Background' 
case) are then tabulated in Table 3. 
If the imager subarrays are used, then the limits shown in 
Table 3 should be increased by a factor equal to 
2.78 seconds divided by the subarray frame time listed in Paper VIII, For 
example, the SUB64 subarray's 0.085 second frame time should allow a 0.4 
Jansky target to be observed using the F560W filter without saturating.

For the LRS, where the PSF is co-added along the dispersion direction, we 
use $f_{br\mathunderscore pix\mathunderscore lrs} =$ 0.36 (the square root 
of the value used for the imager) for $\lambda \le $ 8 $\mu $m and for 
$\lambda $ \textgreater 8 $\mu $m, 
\begin{equation}
{\begin{array}{*{20}c}
f_{br\mathunderscore pix\mathunderscore LRS}=0.36\thinspace \left( 
\raise0.7ex\hbox{${8\thinspace \mu m}$} \!\mathord{\left/ {\vphantom 
{{8\thinspace \mu m} \lambda 
}}\right.\kern-\nulldelimiterspace}\!\lower0.7ex\hbox{$\lambda $} \right)\\
\thinspace \thinspace \thinspace \thinspace \thinspace \thinspace \\
\end{array} }
\end{equation}
\noindent
As for the imager subarrays, if the `SLITLESSPRISM' subarray with its 0.16 
second frame time is used, then the limits presented below should be 
increased by a factor of 17. 

For the MRS, we scale $f_{br\mathunderscore pix\mathunderscore imager}$ by 
the ratio of the solid angle viewed by each MRS pixel to the solid angle 
viewed by an imager pixel. The resulting bright source limits are plotted in 
\ref{subsec:sensitivity}. We note that the nominal MRS readout 
pattern currently uses `SLOW' mode (as defined in Paper VIII). If `FAST' mode 
is not implemented for the MRS on orbit, the MRS bright source limits 
presented here should be reduced by a factor of $\sim$ 10. 

\subsection{Sensitivity}
\label{subsec:sensitivity}
We can now write the formula used to calculate the limiting sensitivity, 
defined as the target flux needed to achieve a S/N ratio of 10 in a 10,000 
second observation. 
\begin{equation}
F_{sens}={10\thinspace 
k}_{margin}k_{extr}k_{ff}\frac{P_{point\mathunderscore tgt}}{\left( 
\raise0.7ex\hbox{$S_{int}$} \!\mathord{\left/ {\vphantom {S_{int} 
N_{int}}}\right.\kern-\nulldelimiterspace}\!\lower0.7ex\hbox{$N_{int}$} 
\right)\sqrt {N_{phot}\frac{t_{eff}}{t_{int}}} }
\end{equation}

\noindent
Here, we introduce the term `$k_{ff}$' to account for the error due to 
differences in the gain of the individual pixels which are not corrected by 
standard division by a reference pixel flat, (and which we refer to as flat 
fielding noise). We calculate its contribution to scale with the total 
signal collected as, 
\begin{equation}
\label{eq6}
k_{ff}={1+K}_{ff}\sqrt {\frac{t_{eff}}{t_{int}}\left( i_{tot}t_{int} 
\right)} 
\end{equation}
The factor `$K_{ff}$' is set equal to 10$^{\mathrm{-3}}$ for the 
spectrometer, 10$^{\mathrm{-5}}$ for the imager at $\lambda $ \textgreater 
12 $\mu $m, and 10$^{\mathrm{-4}}$ in all other cases. 
The factor $k_{extr}$ is used to account for the noise penalty associated 
with extracting source fluxes from data which is not fully (Nyquist) 
sampled. On this basis, for MIRI it is set equal to 1.1 for the F560W imager 
filter and the MRS and equal to 1.0 in all other cases. 
Finally, we retain a factor `$k_{margin}$' ($=$ 1.1) to account for 
unexpected impacts on the delivered sensitivity that may arise before MIRI 
starts operation on orbit. 

Table 3 lists the limiting sensitivities for each 
filter in the imager. For the MRS, \figurename~\ref{fig10} plots the sensitivity for 
the detection of a spectrally unresolved emission line in a spatially 
unresolved target, with units of watt m$^{\mathrm{-2}}$, while 
\figurename~\ref{fig11} plots the sensitivity for a spatially 
unresolved continuum source. The equivalent plots for the LRS are shown in 
Figures 12 and 13.

To allow the MRS sensitivities to be calculated conveniently we 
have made second order polynomial fits to the model predictions within each 
MRS sub-band for the case of unresolved spectral lines and the Case 2 (high) 
radiative background. These fits are of the form

\begin{equation}
\label{eq6}
LF = A x^2 + B x + C
\end{equation}

\noindent
where LF is the minimum line flux (for an unresolved line from a point source) that can be
detected at 10$\sigma$ in 10,000 seconds of integration and $x = \lambda - \lambda_0$.
The fit coefficients are provided in 
Table 4; in general they reproduce the detailed 
model results to within a few percent (a worst case deviation of 13 {\%} is 
seen for Sub-band 3A).

The sensitivity figures for continuum targets can be calculated using the 
spectral resolving powers listed in Paper VI. For extended sources, we note 
that the area of the photometric aperture used in the model is 0.83 
arcsec$^{\mathrm{2}}$ at $\lambda =$ 10 $\mu $m and scales as $\lambda 
^{\mathrm{2}}$. To take a specific example, if we wish to know the MRS 
sensitivity for a spectral line from an extended source at $\lambda$ = 
12.8 $\mu $m, \ref{sec:conclusion} allows us to calculate the 
unresolved sensitivity as 0.86 $\times 10^{\mathrm{-20}}$ Watt 
m$^{\mathrm{-2}}$ . We divide this figure by the aperture area (1.36 
arcsec$^{\mathrm{2}})$ to give a figure of 0.63 $\times$ 
10$^{\mathrm{-20}}$ Watt m$^{\mathrm{-2\thinspace }}$arcsec$^{\mathrm{-2}}$.

\section{CONCLUSION}
\label{sec:conclusion}
The limiting sensitivities presented above describe the MIRI Team's most 
accurate estimate of the performance that will be achieved on orbit, based 
on the best available test results and analysis. We note that the figures 
are not significantly different from expectations presented previously (for 
example in Glasse et al., 2006), with the notable exception of the worse 
than expected performance for Channel 4. The overall effect of updating the 
sensitivity model has been to replace factors assigned to areas of potential 
risk with measured parameters describing measured features of the 
instrument. 

Remaining areas of uncertainty in the model are probably dominated by the 
JWST radiative background at long wavelengths; the high background 
bounding scenario we present here should be conservative. 
Other effects, such as the potential need to 
use sub-array readouts (see Paper VIII) to avoid saturating on the 
background will have a much smaller impact unless the observatory background is 
significantly higher (more than a factor of two) than the higher of the levels assumed here. The accuracy of the sensitivity 
predictions is estimated to be in the region $\pm$20 {\%}, substantially 
less than the error bars on the PCE plots in Figures 2 and 3.

We have confirmed that the performance of MIRI continues to promise to meet 
its ambitious science goals. In conjunction with the bright source limits, 
these revised sensitivity figures can be regarded as accurate enough to 
allow potential MIRI observers some confidence in starting to refine and 
focus their planned observing programmes.

\section{Acknowledgments}
The work presented is the effort of the entire MIRI team and the enthusiasm within the MIRI partnership is a significant factor in its success. MIRI draws on the scientific and technical expertise many organizations, as summarized in Papers I and II. 
We also thank Jane Rigby and Paul Lightsey for helpful comments. 

We would like to thank the following National and International
Funding Agencies for their support of the MIRI development: NASA; ESA;
Belgian Science Policy Office; Centre Nationale D'Etudes Spatiales;
Danish National Space Centre; Deutsches Zentrum fur Luft-und Raumfahrt
(DLR); Enterprise Ireland; Ministerio De Economi{\'a} y Competividad;
Netherlands Research School for Astronomy (NOVA); Netherlands
Organisation for Scientific Research (NWO); Science and Technology Facilities
Council; Swiss Space Office; Swedish National Space Board; UK Space
Agency.

\clearpage

\begin{deluxetable}{ccc}
\tabletypesize{\footnotesize}
\tablecolumns{3}
\tablewidth{0pt}
\tablecaption{Grey-body components of the background model used for sensitivity calculations.  Components A and B 
are fits to the scattered and emissive components of the zodiacal dust spectrum while C to F represent the observatory 
straylight.  See the text for a more detailed explanation of the terms.}
\tablehead{\colhead{Component}             &
	\colhead{Emissivity}       &       
          \colhead{Temperature(K)}   \\
 	 }
\startdata
A  & $4.20 \times 10^{-14}$  &  5500 \\
B  &  $4.30 \times 10^{-6}$      &  270 \\
C  & $3.35 \times 10^{-7}$     &  133.8 \\
D  & $9.70 \times 10^{-5}$     &   71.0  \\
E  & $1.72 \times 10^{-3}$     &  62.0   \\
F  & $1.48 \times 10^{-2}$     &  51.7  \\   
(G)  &  $1.31 \times 10^{-4}$ &  86.7 \\
\enddata
\label{tab1}
\end{deluxetable}

\clearpage

\begin{deluxetable}{ccc}
\tabletypesize{\footnotesize}
\tablecolumns{3}
\tablewidth{0pt}
\tablecaption{Effective transmission due to diffractive losses at the short and long wavelength limits of the four MRS channels}
\tablehead{\colhead{Channel}             &
	\colhead{$\tau_{MRS\_diff}$}       &       
          \colhead{$\tau_{MRS\_diff}$}   \\
          \colhead{}                                    &
          \colhead{at $\lambda_{short}$} &
          \colhead{at $\lambda_{long}$}  \\
 	 }
\startdata
1  & 0.95  &  0.91 \\
2  & 0.94  &  0.91 \\
3  & 0.92  &  0.87 \\
4  & 0.92  &  0.86  \\
\enddata
\label{tab2}
\end{deluxetable}

\clearpage

\begin{deluxetable}{ccccc}
\tabletypesize{\footnotesize}
\tablecolumns{5}
\tablewidth{0pt}
\tablecaption{Limiting faint source detection limits (10$\sigma$ in 10,000 seconds) and bright source limits for the MIRI imager. Note that the detection limits are quoted in micro-Jansky and the bright source limits (which are calculated for the 'high-background' case only) are quoted in milli-Jansky and Jansky.}
\tablehead{\colhead{Filter}             &
	\colhead{Low Background}       &       
          \colhead{High Background}   &
          \colhead{Brt. Src. Limit (mJy)}        &
          \colhead{Brt. Src. Limit (Jy)}          \\
          \colhead{}                                    &
          \colhead{detection limit ($\mu$Jy)} &
          \colhead{detection limit ($\mu$Jy)}  &
          \colhead{full frame}                           &
          \colhead{64 $\times$ 64 subarray}  \\
 	 }
\startdata
F560W  & 0.16  &  0.16  &  13  &  0.42  \\
F770W  & 0.26  &  0.27  &  7.4 &  0.24  \\
F1000W  & 0.58  &  0.59  &  16  &  0.52  \\
F1130W  & 1.41  &  1.50  &  69  &  2.25  \\
F1280W  & 0.94  &  1.12  &  29  &  0.95  \\
F1500W  & 1.48  &  2.06  &  37  &  1.23  \\
F1800W  & 3.65  &  5.15  &  66  &  2.2   \\
F2100W  & 7.48  &  9.66 & 66  &  2.2  \\
F2550W  & 27.2  &  31.9  &  192 & 6.4  \\
\enddata
\end{deluxetable}

\clearpage

\begin{deluxetable}{ccccc}
\tabletypesize{\footnotesize}
\tablecolumns{5}
\tablewidth{0pt}
\tablecaption{Coefficients of second order polynomial fits to the model MRS spectral line sensitivities (the noise equivalent line intensity to achieve a signal to noise ratio of 10 in a 10,000 second observation), for Case 2 (high) radiative background.  The spectral coverages of the individual MRS sub-bands are provided in Paper VI.}
\tablehead{\colhead{MRS Channel}             &
	\colhead{$\lambda_0$}       &       
          \colhead{A}   &
          \colhead{B}        &
          \colhead{C}          \\
          \colhead{and Sub-band}                                    &
          \colhead{$\mu m$}                                           &
          \colhead{$\times 10^{-20}$ W m$^{-2}$ $\mu m^{-2}$} &
          \colhead{$\times 10^{-20}$ W m$^{-2}$ $\mu m^{-1}$}  &
          \colhead{$\times 10^{-20}$ W m$^{-2}$ }  \\
 }
\startdata
1A  & 5.4  &  0.0074  & -0.2813  &  0.7958  \\
1B  & 6.2  &  0.0859  &  -0.1035 &  0.6206  \\
1C  & 7.2  &  -0.0521 &  -0.0071 &  0.5768  \\
2A  & 8.2  &  0.1004  &  -0.0354 &  0.5376  \\
2B  & 9.5  &  -0.0046 &  0.0616  &  0.5729  \\
2C & 10.9 &  0.0180  &  0.0373  &  0.6061  \\
3A & 12.6 &  0.0719  &  0.0862  &  0.8970  \\
3B & 14.5 &  0.0961 & 0.0329  &  1.0170  \\
3C & 16.8 &  0.0205 & 0.2122 & 1.4381  \\
4A & 19.4 &  0.0392 & 1.2568 &  6.7650  \\
4B & 22.6  & 0.4080  & 2.6966  & 11.8131  \\
4C($\lambda < 27.5 \mu m$) & 25.5  & 1.7863 & 7.1873 &  21.7046  \\
4C($\lambda \ge 27.5 \mu m$) & 27.9  & 21.1600 &  42.3280 & 58.7956  \\
\enddata
\end{deluxetable}

\clearpage

\clearpage

\clearpage

\clearpage

\begin{figure}[htbp]
\centerline{\includegraphics[width=5.0in]{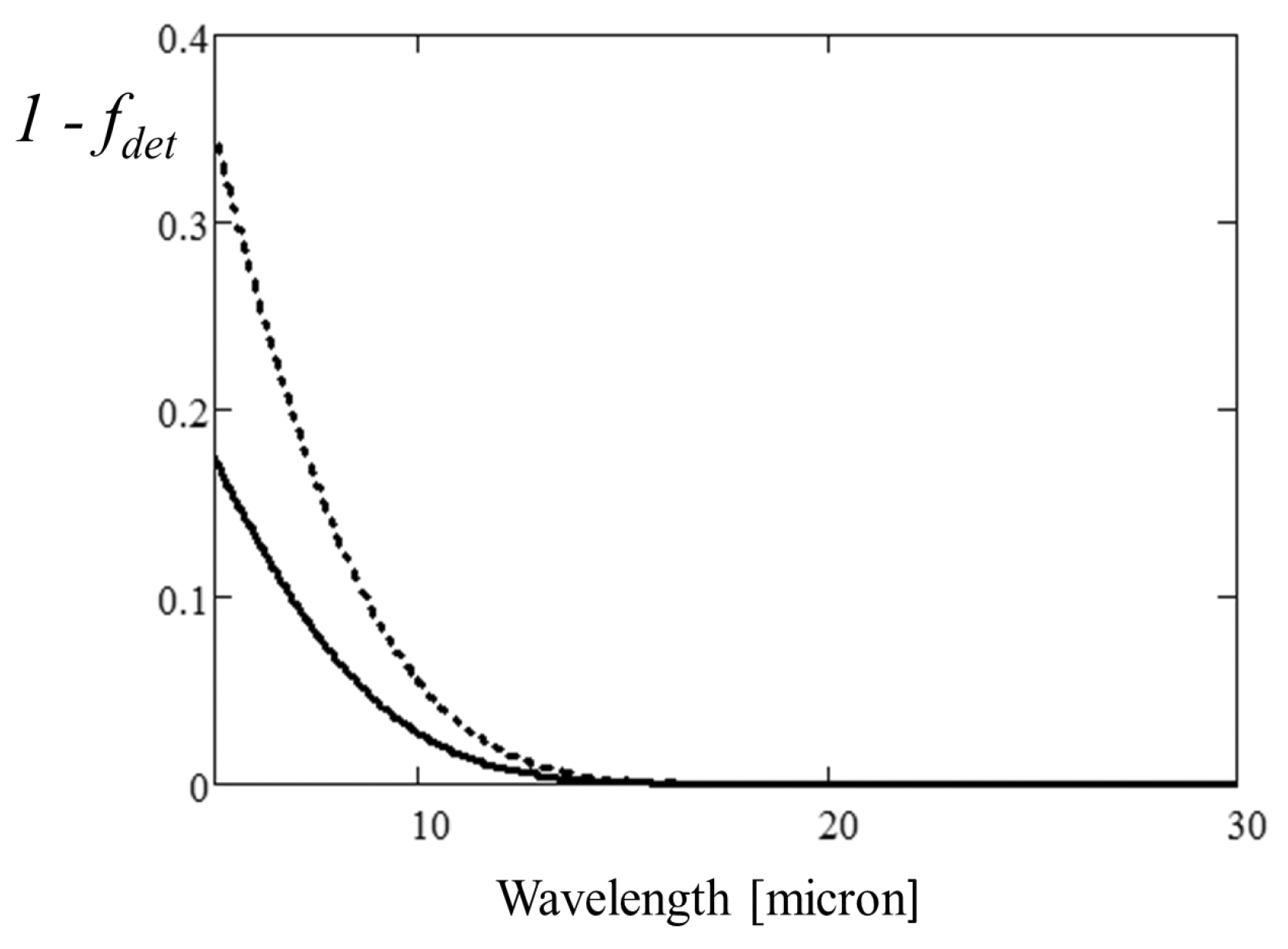}}
\caption{The fraction of light lost from the photometric aperture due to detector scattering as a function of wavelength:  The imager is shown as a solid line) and the MRS as a dashed line. }
\label{fig4}
\end{figure}

\clearpage

\begin{figure}[htbp]
\centerline{\includegraphics[width=5.0in]{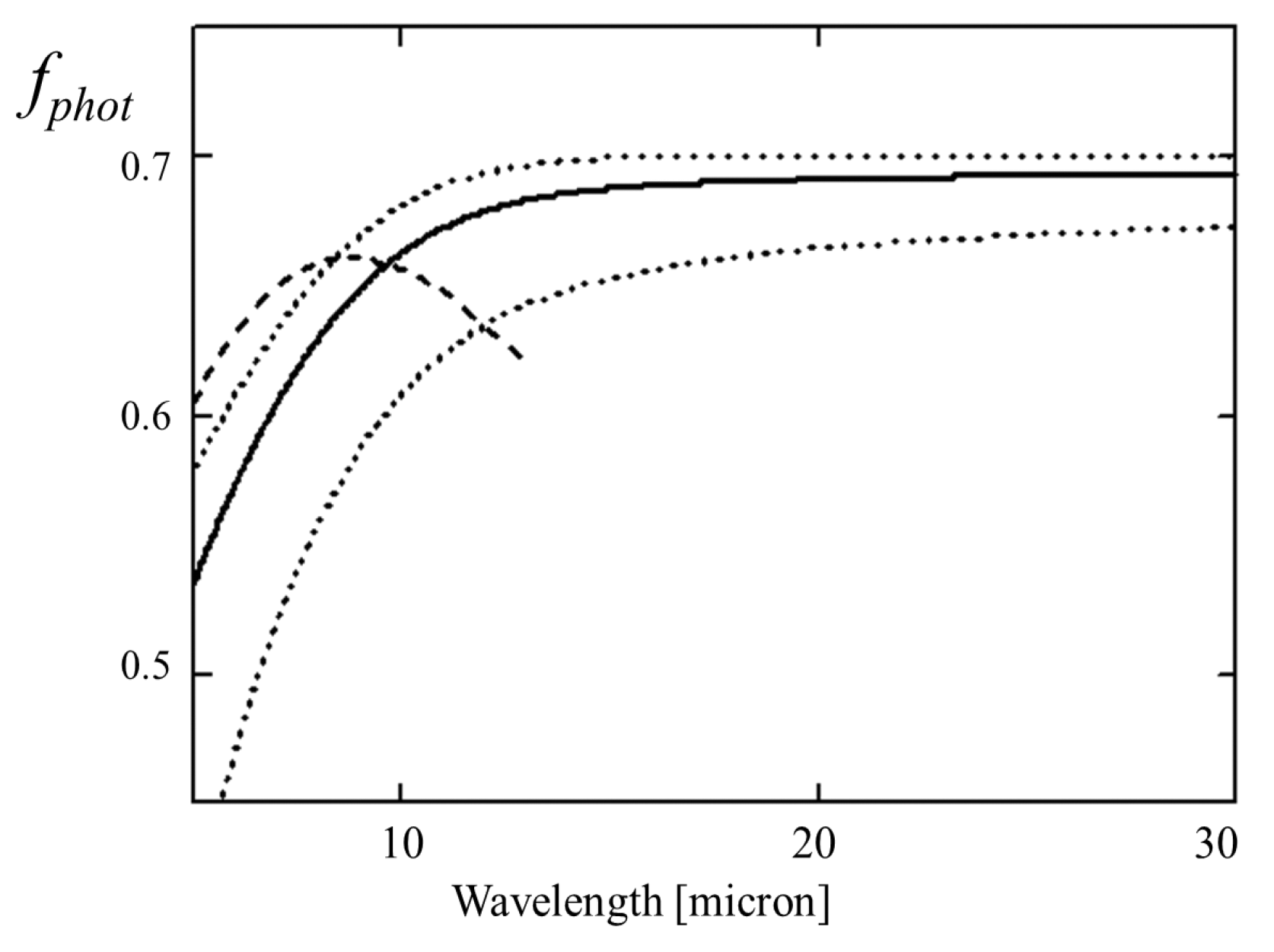}}
\caption{Fraction of point source energy falling within the photometric aperture.  The nominal case for the imager is shown as a solid line, and for the LRS as a dashed line.  The dotted lines show the variations seen for the imager under the best and worst bounding cases of image quality. }
\label{fig5}
\end{figure}

\clearpage

\begin{figure}[htbp]
\centerline{\includegraphics[width=5.0in]{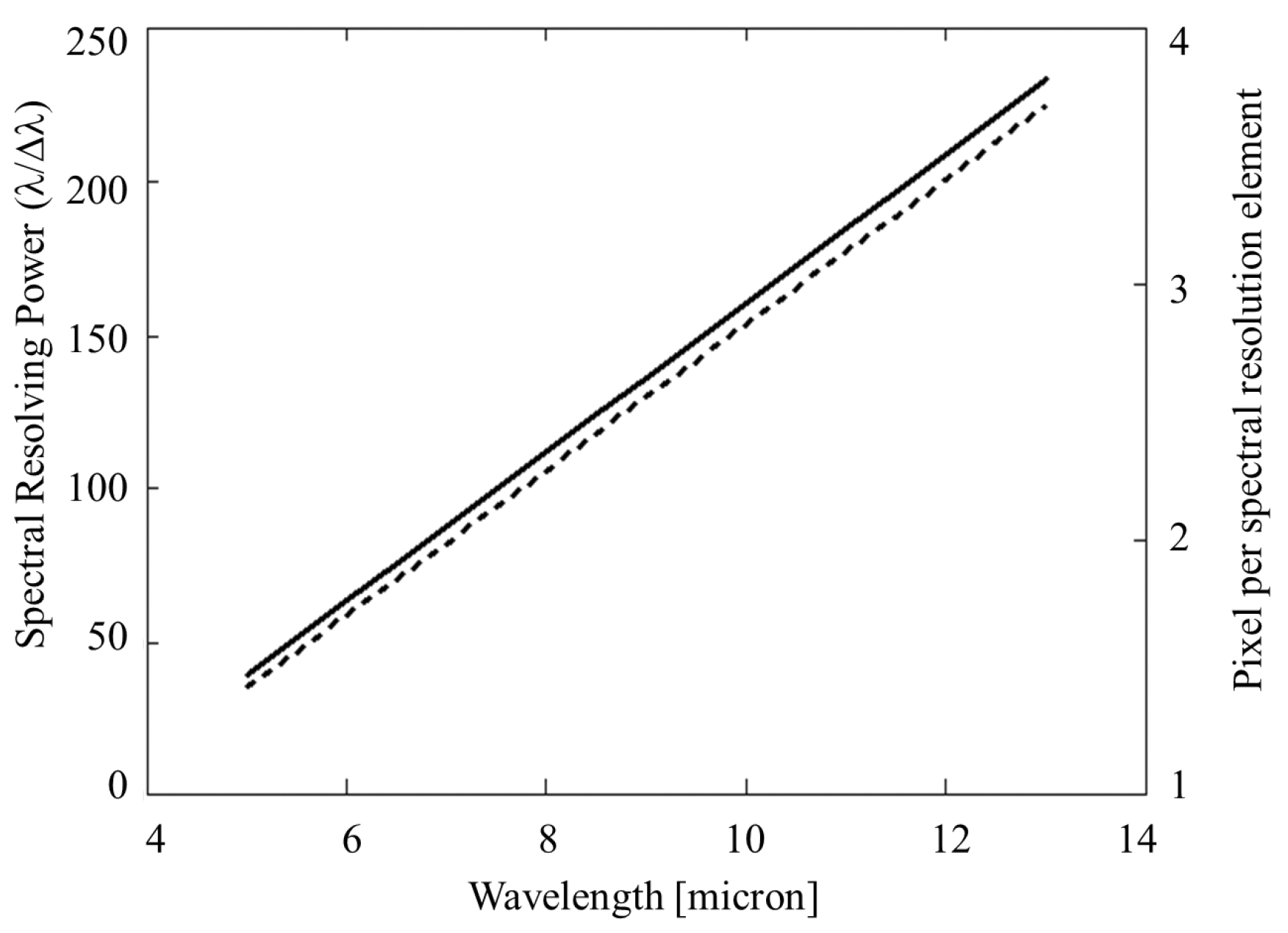}}
\caption{LRS spectral sampling parameters used in the sensitivity model.  The spectral resolving power (solid line) and number of pixels per spectral resolution element (dashed line) are shown as a function of wavelength.}
\label{fig6}
\end{figure}

\clearpage

\begin{figure}[htbp]
\centerline{\includegraphics[width=5.0in]{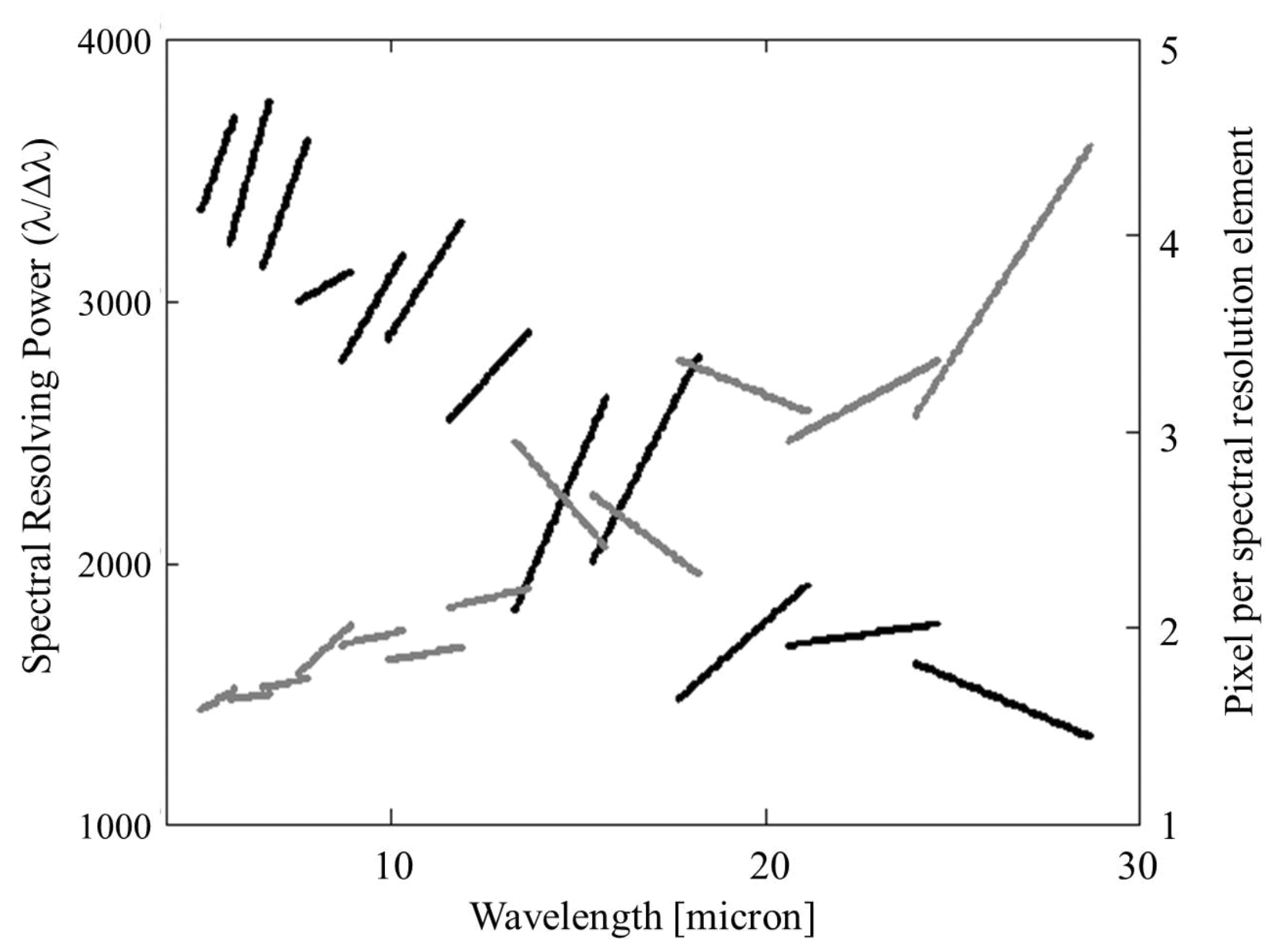}}
\caption{MRS spectral sampling parameters used in the sensitivity model.    The spectral resolving power (black line) and number of pixels per spectral resolution element (grey lines) are shown as a function of wavelength.}
\label{fig7}
\end{figure}

\clearpage

\begin{figure}[htbp]
\centerline{\includegraphics[width=5.0in]{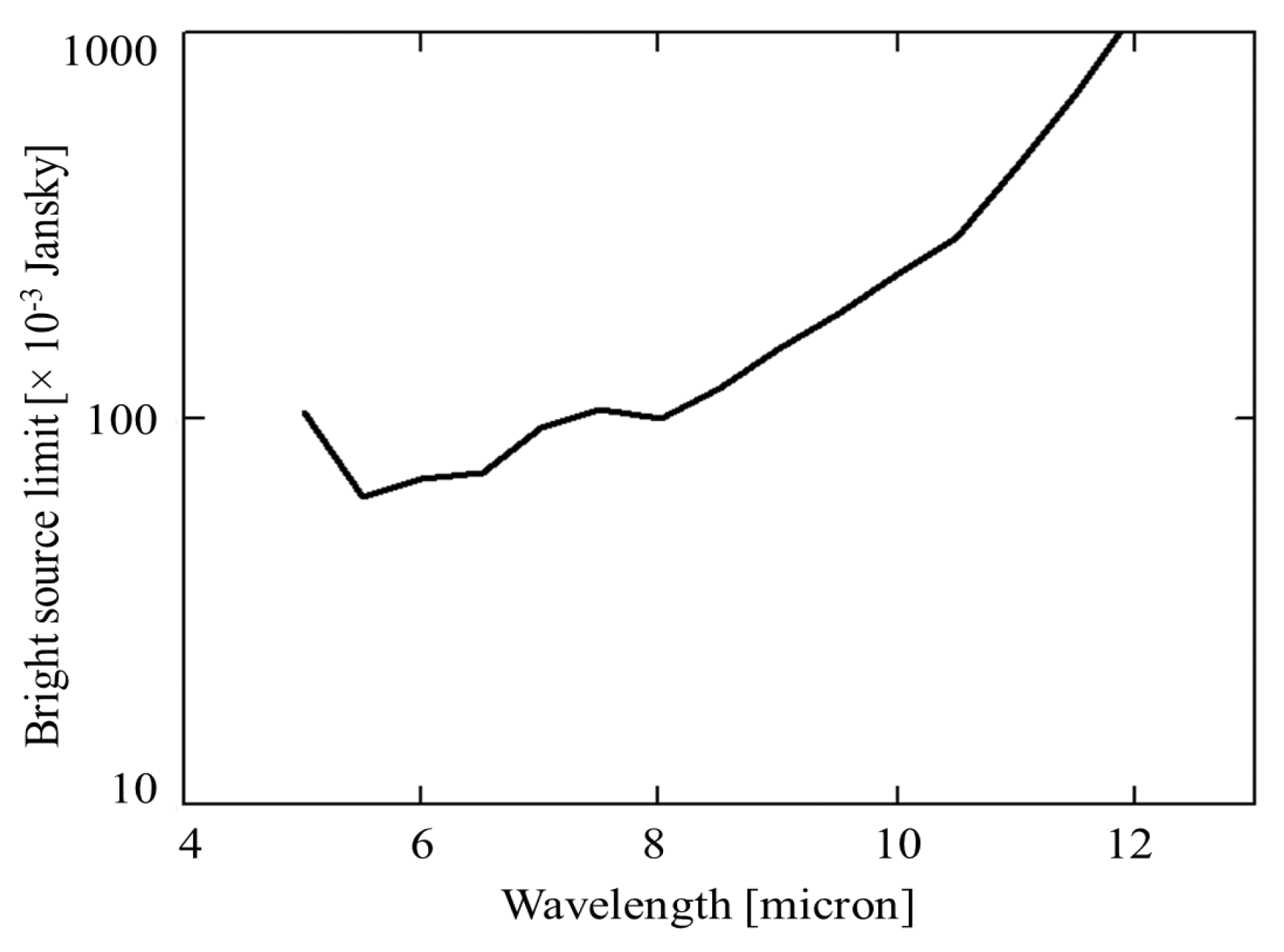}}
\caption{Bright source limits (calculated for the ‘high background’ case only) for the MIRI Low Resolution Spectrometer.  The minimum value is 63 milli-Jansky at $\lambda$ = 5.5 $\mu$m.}
\label{fig8}
\end{figure}

\clearpage

\begin{figure}[htbp]
\centerline{\includegraphics[width=5.0in]{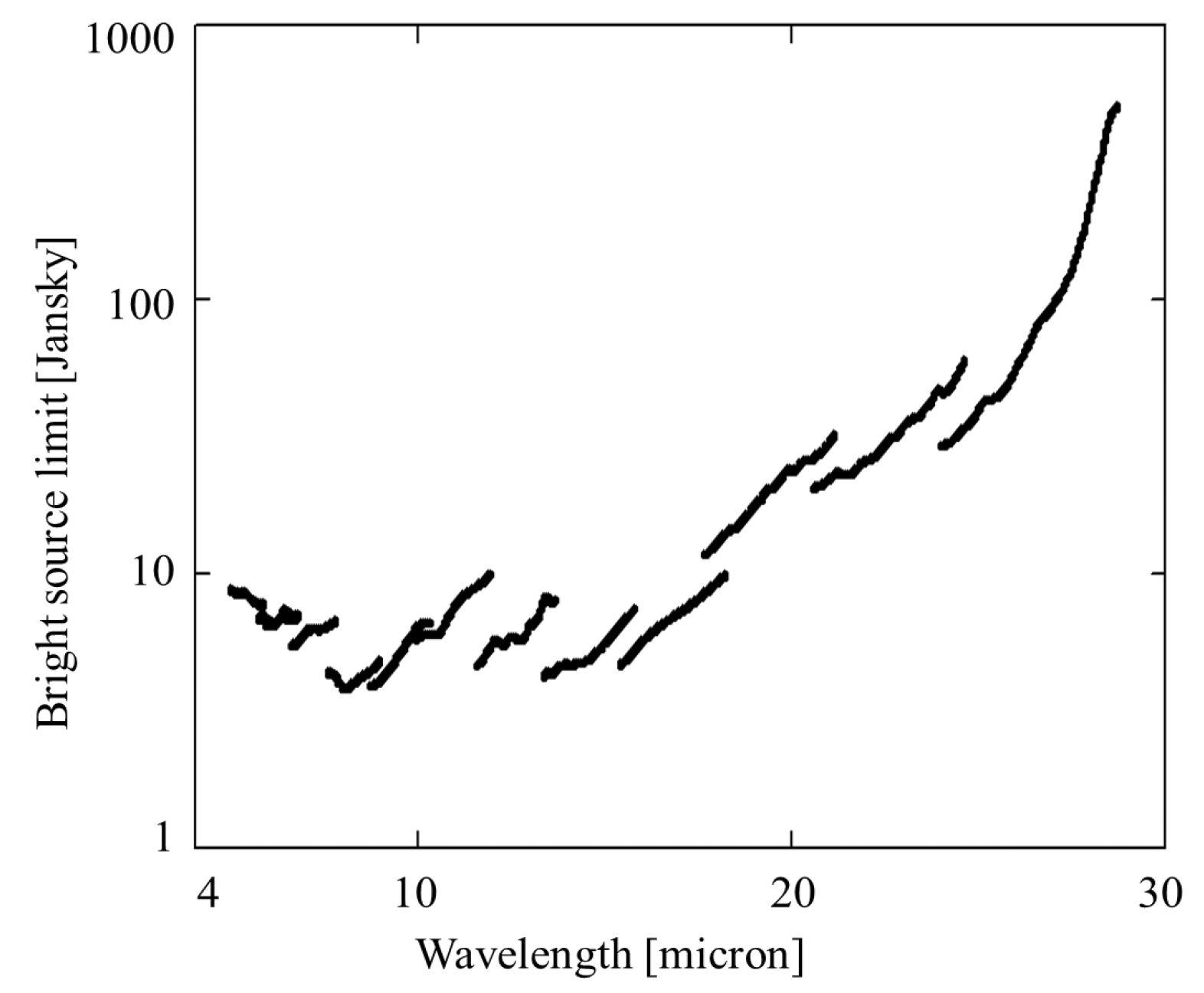}}
\caption{Bright source limits (calculated for the ‘high background’ case only) for the MIRI Medium Resolution Integral Field Spectrometer.  The minimum value is 3.8 Jansky at $\lambda$ = 8 $\mu$m.}
\label{fig9}
\end{figure}

\clearpage

\begin{figure}[htbp]
\centerline{\includegraphics[width=5.0in]{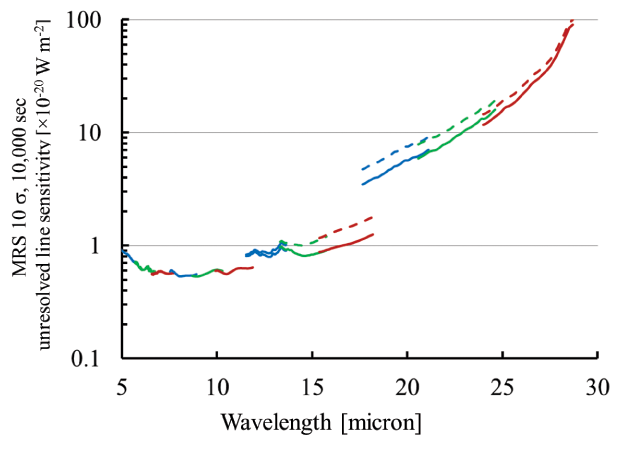}}
\caption{MRS limiting sensitivity for the detection of an unresolved spectral line in a spatially unresolved target (units of 10$^{-20}$ Watt m$^{-2}$). The dashed lines are for the high background case.}
\label{fig10}
\end{figure}

\clearpage

\begin{figure}[htbp]
\centerline{\includegraphics[width=5.0in]{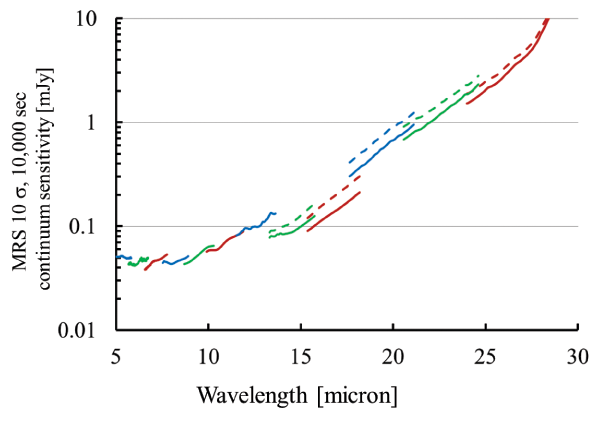}}
\caption{MRS limiting sensitivity for the detection of the continuum spectrum for a spatially unresolved target (units of milli-Jansky). 
The dashed lines are for the high background case.}
\label{fig11}
\end{figure}

\clearpage

\begin{figure}[htbp]
\centerline{\includegraphics[width=5.0in]{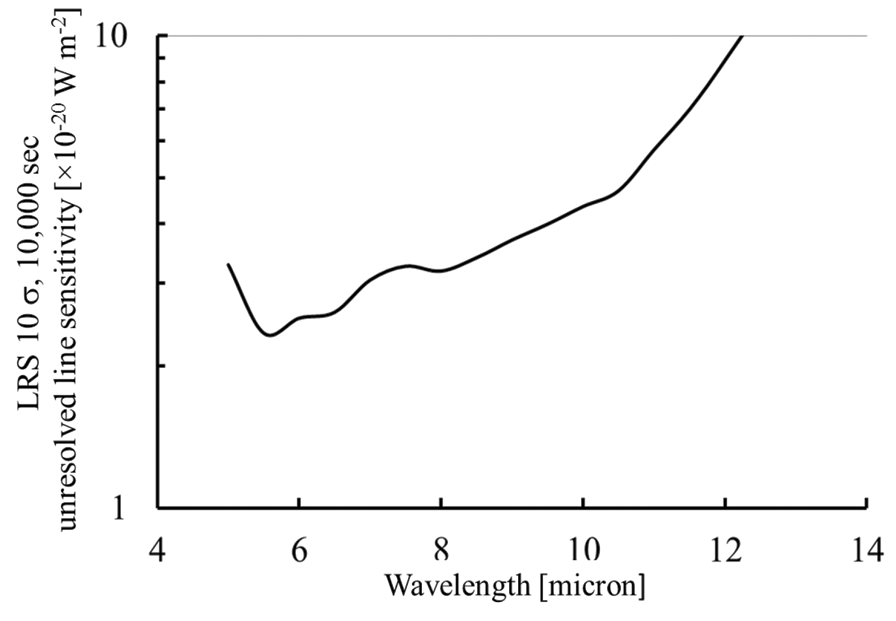}}
\caption{LRS limiting sensitivity for the detection of a spectrally and spatially unresolved target (units of 10$^{-20}$ Watt m$^{-2}$).}
\label{fig12}
\end{figure}

\clearpage

\begin{figure}[htbp]
\centerline{\includegraphics[width=5.0in]{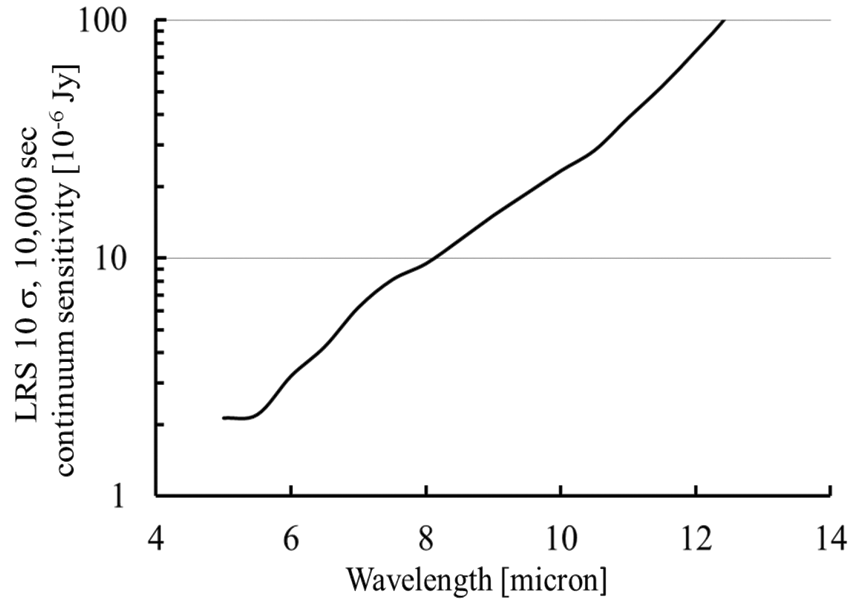}}
\caption{LRS limiting sensitivity for the detection of the continuum spectrum for a spatially unresolved target (units of microJansky).}
\label{fig13}
\end{figure}

\end{document}